\def\nk{n_{\rm b}}

\def\Pb{P_{\rm b}}

\def\rfr#1{Equation~(\ref{#1})}
\def\rfrs#1#2{Eqs.~(\ref{#1})~to~(\ref{#2})}
\def\Rfr#1{Eq. (\ref{#1})}
\def\Rfrs#1#2{Eq. (\ref{#1})-eq. (\ref{#2})}

\def\virg#1{``#1"}

\def\eqi{\begin{equation}}
\def\eqf{\end{equation}}
\def\eqia{\begin{eqnarray}}
\def\eqfa{\end{eqnarray}}

\def\rp#1#2{{#1\over#2}}
\def\lb#1{\label{#1}}

\def\bds#1{\boldsymbol{#1}}

\def\ton#1{\left(#1\right)}
\def\qua#1{\left[#1\right]}
\def\grf#1{\left\{#1\right\}}

\documentclass[onecolumn]{aastex}
\usepackage{morefloats}
\usepackage[title]{appendix}
\usepackage{hyperref}
\usepackage{booktabs}
\usepackage[table,xcdraw]{xcolor}
\usepackage{multirow}
\usepackage{rotating,tabularx}
\usepackage{float}
\usepackage{enumerate}
\usepackage{rotating}
\usepackage[polutonikogreek,english]{babel}
\usepackage{amsmath,starfont,textgreek,w-greek,wasysym}
\usepackage{amsthm}
\usepackage{amscd,lineno}
\usepackage{amssymb,dsfont}
\usepackage{graphicx,epsfig}
\usepackage{txfonts}
\usepackage{xr-hyper}

\RequirePackage{color}

\newcommand{\emaila}{lorenzo.iorio@libero.it}

\allowdisplaybreaks[1]

\begin{document}

\title{New general relativistic contributions to Mercury's orbital elements and their measurability}

\shortauthors{L. Iorio}

\author{Lorenzo Iorio\altaffilmark{1} }
\affil{Ministero dell'Istruzione, dell'Universit\`{a} e della Ricerca
(M.I.U.R.)-Istruzione
\\ Permanent address for correspondence: Viale Unit\`{a} di Italia 68, 70125, Bari (BA),
Italy}

\email{\emaila}

\begin{abstract}
We numerically and analytically work out the first-order post-Newtonian (1pN) orbital effects induced on the semimajor axis $a$, the eccentricity $e$, the inclination $I$, the longitude of the ascending node $\Omega$, the longitude of perihelion $\varpi$, and the mean longitude at epoch $\epsilon$ of a test particle orbiting its primary, assumed static and spherically symmetric, by a distant massive third body X. For Mercury, the rates of change of the linear trends found are $\dot I_\mathrm{1pN}^\mathrm{X} = -4.3\,\mathrm{microarcseconds\,per\,century}\,\ton{\mu\mathrm{as\,cty}^{-1}}$,  $\dot\Omega_\mathrm{1pN}^\mathrm{X} = 18.2\,\mu\mathrm{as\,cty}^{-1}$,  $\dot\varpi_\mathrm{1pN}^\mathrm{X} = 30.4\,\mu\mathrm{as\,cty}^{-1}$,  $\dot\epsilon_\mathrm{1pN}^\mathrm{X} = 271.4\,\mu\mathrm{as\,cty}^{-1}$, respectively. Such values, which are due to the added actions of the other planets from Venus to Saturn, are essentially at the same level of, or larger by one order of magnitude than, the latest formal errors in the Hermean orbital precessions  calculated with the EPM2017 ephemerides. The perihelion precession $\dot\varpi_\mathrm{1pN}^\mathrm{X}$ turns out to be smaller than some values recently appeared in the literature in view of a possible measurement with the ongoing BepiColombo mission. Linear combinations of the supplementary advances of the Keplerian orbital elements for several planets, if determined experimentally by the astronomers, could be set up in order to disentangle the 1pN $N$-body effects of interest from the competing larger precessions like those due to the Sun's quadrupole moment $J_2$ and angular momentum $\bds{S}$.
\end{abstract}

keywords{
gravitation $-$ celestial mechanics $-$ ephemerides $-$ methods: miscellaneous
}

\section{Introduction}\lb{intro}
In its weak-field and slow-motion approximation, general relativity\footnote{See, e.g., \citet{2016Univ....2...23D} and references therein for a recent overview on its status and challenges.} predicts that, in addition to the time-honored first-order post-Newtonian (1pN) gravitoelectric and gravitomagnetic   precessions induced by the mass monopole $M$ (Schwarzschild) and the spin dipole  $\bds{S}$  (Lense-Thirring) moments  of the central body acting as source of the gravitational field, further 1pN orbital effects due to the presence of other interacting masses arise as well \citep{2018PhRvL.120s1101W}.
Let us consider a nonrotating primary of mass $M$, assumed as origin of a locally inertial coordinate system, orbited by a test particle located at $\bds{r}$ and moving with velocity $\bds{v}$. If a distant, pointlike  body X of mass $M_\mathrm{X}$ is present at $\bds{r}_\mathrm{X}$ and moves with velocity
$\bds{v}_\mathrm{X}$ with respect to $M$, the test particle experiences certain 1pN accelerations which, from Eq.\,(4) of \citet{2018PhRvL.120s1101W}, are
\begin{align}
\bds{A}_{G^2} \lb{AG2} &= \rp{2G^2 M M_\mathrm{X}}{c^2 r_\mathrm{X}^3}\qua{ \bds{\hat{r}}  - 6 \ton{\bds{\hat{r}} \cdot {\bds{\hat{r}}}_\mathrm{X}  }{\bds{\hat{r}}}_\mathrm{X} + 3\ton{\bds{\hat{r}} \cdot{\bds{\hat{r}}}_\mathrm{X} }^2  \bds{\hat{r}}  }, \\ \nonumber \\
\bds{A}_{G} \nonumber &= \rp{GM_\mathrm{X} r}{c^2 r^3_\mathrm{X}}\grf{4\,\bds{v}\qua{\ton{\bds{v}\cdot\bds{\hat{r}} } - 3\ton{ \bds{\hat{r}}\cdot{\bds{\hat{r}}}_\mathrm{X} }\ton{\bds{v}\cdot{\bds{\hat{r}}}_\mathrm{X} }  } - \right.\\ \nonumber \\
\lb{AG1}&\left. - v^2\qua{\bds{\hat{r}} -3\ton{\bds{\hat{r}}\cdot{\bds{\hat{r}}}_\mathrm{X}}{\bds{\hat{r}}}_\mathrm{X}  } }, \\ \nonumber \\
\bds{A}_{v_\mathrm{X}} \lb{AGM} &= -\rp{GM_\mathrm{X}}{c^2 r_\mathrm{X}^2}\qua{4\,\bds{v}\times\ton{ {\bds{\hat{r}}}_\mathrm{X}\times{\bds{v}}_\mathrm{X} } - 3\ton{ {\bds{\hat{r}}}_\mathrm{X}\cdot{\bds{v}}_\mathrm{X} }\bds{v}}.
\end{align}
In \rfrs{AG2}{AGM}, which are a particular case of the full 1pN  equations of motion for a  system of $N$ pontlike, massive bodies mutually interacting through gravitation\footnote{See also \citet[Eq.\,(7.11),\,Eq.\,(7.12),\,Eq.\,(8.18)]{1989NCimB.103...63B} with the replacements Earth$\rightarrow$Sun, Sun$\rightarrow$Jupiter, and satellite$\rightarrow$Mercury.} \citep[Eq.\,(9.127)]{2014grav.book.....P}, $G$ is the Newton's gravitational constant, and $c$ is the speed of light in vacuum.

\citet{2018PhRvL.120s1101W} looked at the longitude of perihelion $\varpi$ of Mercury finding an additional contribution to its 1pN secular precession of about
\begin{align}
\dot\varpi_\mathrm{1pN}^\mathrm{X} \nonumber &= 0.22\,\mathrm{milliarcseconds\,per\,century}\,\ton{\mathrm{mas\,cty}^{-1}} = \\ \nonumber \\
& = 220\,\mathrm{microarcseconds\,per\,century}\,\ton{\mu\mathrm{as\,cty}^{-1}}\lb{willo}.
\end{align}
\Rfr{willo} was obtained by making some simplifying assumptions about the orbital geometries of both the perturbed and the perturbing bodies, and includes the combined actions of Venus, Earth, Mars, Jupiter and Saturn. It should be a direct effect of the accelerations of \rfrs{AG2}{AGM}, and an indirect consequence of the interplay between the usual Newtonian $N-$body pull by the other planets and the Sun-only 1pN gravitoelectric acceleration. \rfrs{AG2}{AGM} and all the standard Newtonian and 1pN $N$-body dynamics is routinely modeled in the data reduction softwares of the teams of astronomers producing the planetary ephemerides like the Development Ephemeris (DE)  by the NASA Jet Propulsion Laboratory (JPL) in Pasadena \citep{2014IPNPR.196C...1F}, the Int\'{e}grateur Num\'{e}rique Plan\'{e}taire de l'Observatoire de Paris (INPOP) by the Institut de M\'{e}canique C\'{e}leste et de Calcul des \'{E}ph\'{e}m\'{e}rides (IMCCE) at the Paris Observatory \citep{2018MNRAS.476.1877V}, and the Ephemeris of Planets and the Moon (EPM) by the Institute of Applied Astronomy (IAA) of the Russian Academy of Sciences (RAS) in Saint Petersburg \citep{2015HiA....16..221P}.
\citet{2018PhRvL.120s1101W} claimed that \rfr{willo} would likely be detectable with the ongoing BepiColombo mission to Mercury. According to \citet{2018PhRvL.120s1101W}, it would be so because the expected $\simeq 10^{-6}$ accuracy with which the parameterized Post-Newtonian (PPN) parameters $\beta,\,\gamma$ should be measured by such a spacecraft would correspond to an uncertainty in the main contribution to the Mercury's 1pN perihelion precession $\dot\varpi_\mathrm{1pN}=42.98\,\mathrm{arcseconds\,per\,century}\,\ton{\arcsec\,\mathrm{cty}^{-1}}$ as little as
\eqi
\delta\dot\varpi_\mathrm{1pN}\simeq 0.03\,\mathrm{mas\,cty}^{-1}=30\,\mu\mathrm{as\,cty}^{-1}.
\eqf

\citet{2018EPJC...78..549I}, after having pointed out that the indirect, mixed\footnote{To avoid possible misunderstanding, we clarify that \rfrs{AG2}{AGM} are dubbed as \virg{cross-terms} by \citet{2018PhRvL.120s1101W}, while here such a definition designates the interplay among the standard Newtonian $N$-body and 1pN Sun's monopole accelerations.} effects should likely be not measurable in practical planetary data reductions,
analytically worked out the direct perihelion precessions due to \rfrs{AG2}{AGM}  for arbitrary orbital configurations of both the test particle and the perturbing body X. The total 1pN rate of change induced on the perihelion of Mercury by all the other planets of the solar system from Venus to Saturn would amount to \citep[Table\,2]{2018EPJC...78..549I}
\eqi
\dot\varpi_\mathrm{1pN}^\mathrm{X} =0.15\,\mathrm{mas\,cty}^{-1} = 150\,\mu\mathrm{as\,cty}^{-1}.\lb{mini}
\eqf
\citet{2018EPJC...78..549I} showed also that \rfr{mini} would  likely be overwhelmed by the larger systematic errors due to the mismodeling in the competing secular precessions due to the Sun's oblateness $J_2$ and angular momentum $\bds S$ (1pN Lense-Thirring effect).

In this paper, we will show that the value reported in \rfr{mini} is, in fact, wrong because of an error by \citet{2018EPJC...78..549I} in the calculation of the precession due to \rfr{AG1}. The correct size of the overall 1pN $N-$body perihelion precession of Mercury will turn out to be even smaller than \rfr{mini}, thus enforcing the pessimistic conclusions of \citet{2018EPJC...78..549I} about its possible measurability. As such, we will further explore the consequences of \rfrs{AG2}{AGM} by numerically working out the secular shifts induced by them on all the other orbital elements, i.e. the semimajor axis $a$, the eccentricity $e$, the inclination $I$, the longitude of the ascending node $\Omega$, and the mean longitude at epoch $\epsilon$, and will compare them with the uncertainties in the planetary orbital motions inferred by \citet{2019AJ....157..220I} from the most recent version of the EPM ephemerides \citep{2018AstL...44..554P}. Indeed, if and when the astronomers will observationally produce the supplementary rates of change $\Delta\dot a_\mathrm{obs},\,\Delta\dot e_\mathrm{obs},\,\Delta\dot I_\mathrm{obs},\,\Delta\dot\Omega_\mathrm{obs},\,\Delta\dot\varpi_\mathrm{obs}$, and $\Delta\dot\epsilon_\mathrm{obs}$ of as many planets as possible, it will be possible to generalize the approach proposed by\footnote{At that time, the aliasing Newtonian effect which should have been disentangled from the Sun-only 1pN gravitoelectric perihelion precession by looking at other planets or highly eccentric asteroids was due to the solar quadrupole mass moment $J_2$.}  \citet{1990grg..conf..313S} by suitably combining them in order to disentangle the effects of \rfrs{AG2}{AGM} in from the other competing precessions due to, e.g., the Sun's $J_2$ and $\bds S$.
\section{The 1pN $N-$body secular changes of the orbital elements}\lb{calcoli}
\subsection{Numerical integration of the equations of motion}
\begin{figure*}[ht]
\centering
\begin{tabular}{cc}
\includegraphics[width=7cm]{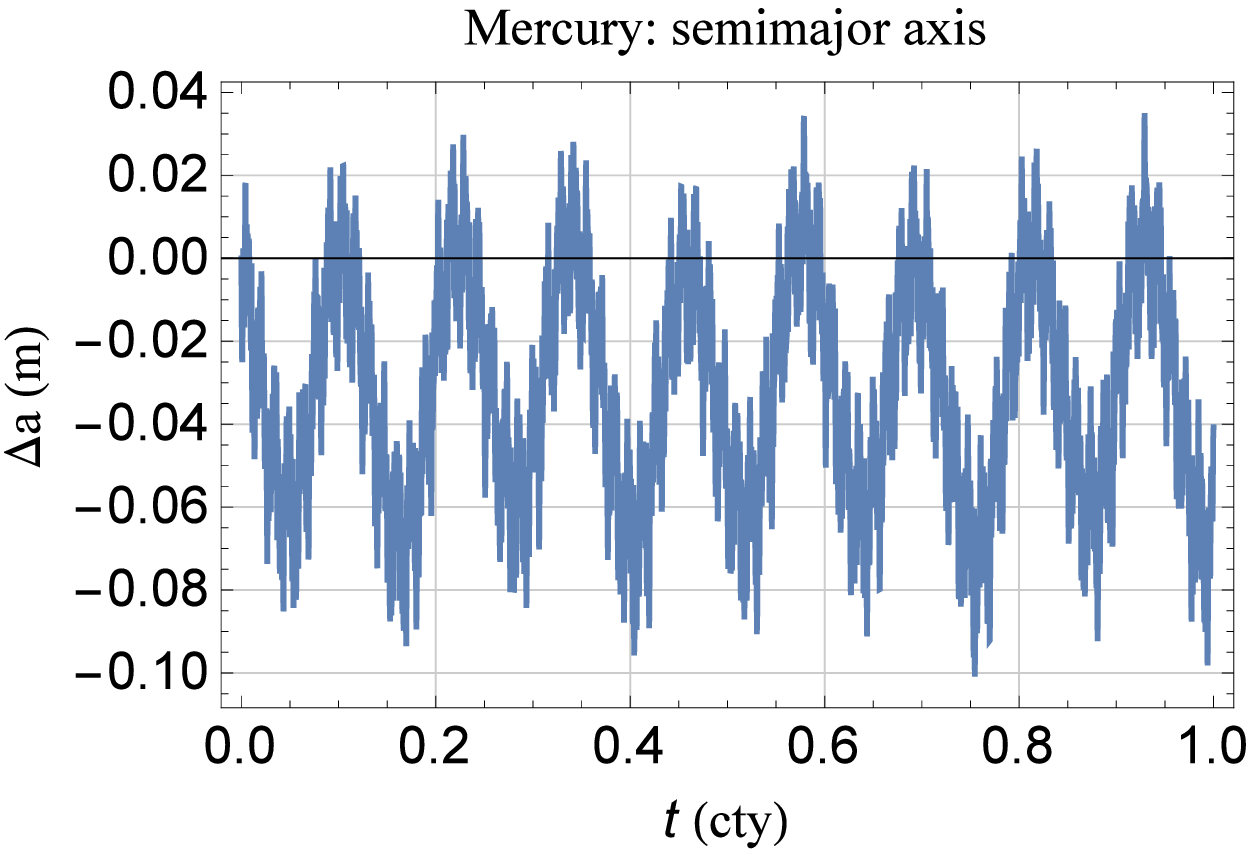}&\includegraphics[width=7cm]{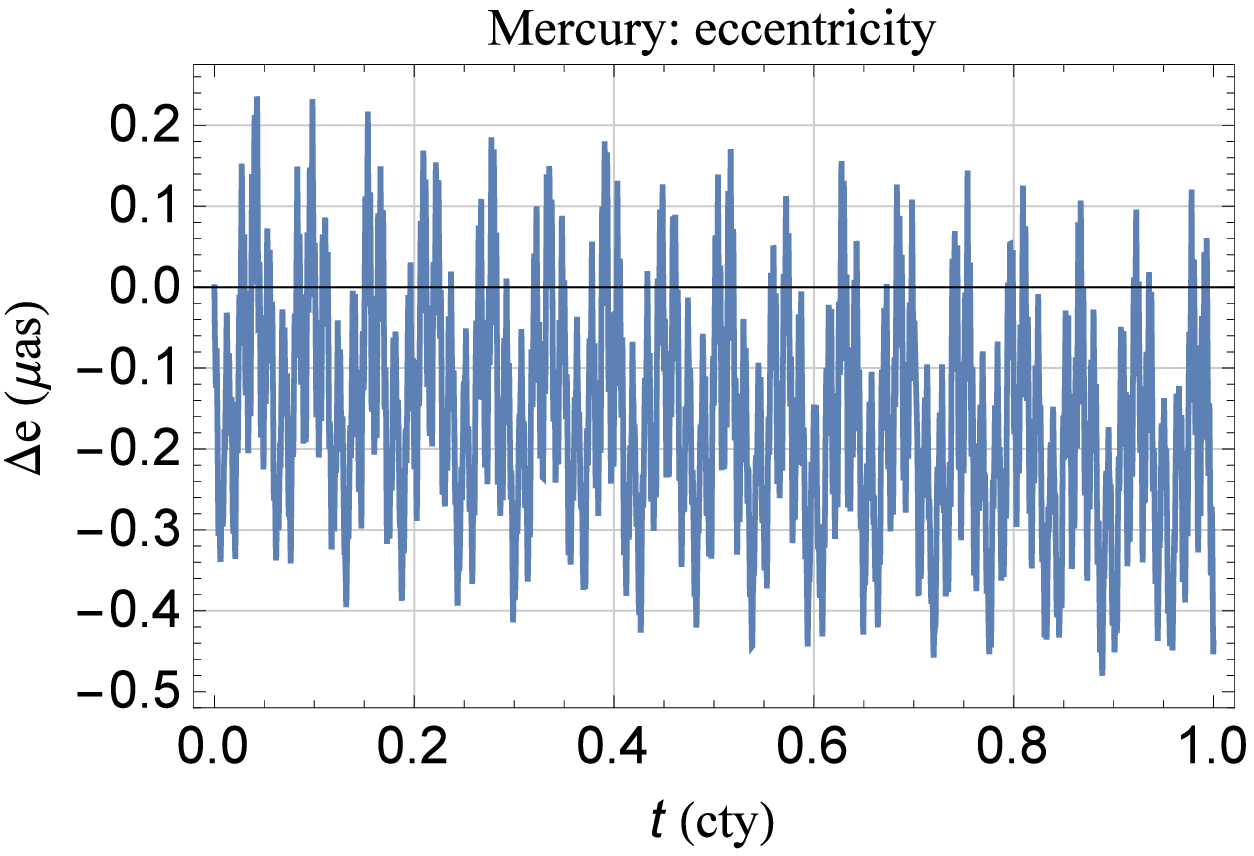}\\
\includegraphics[width=7cm]{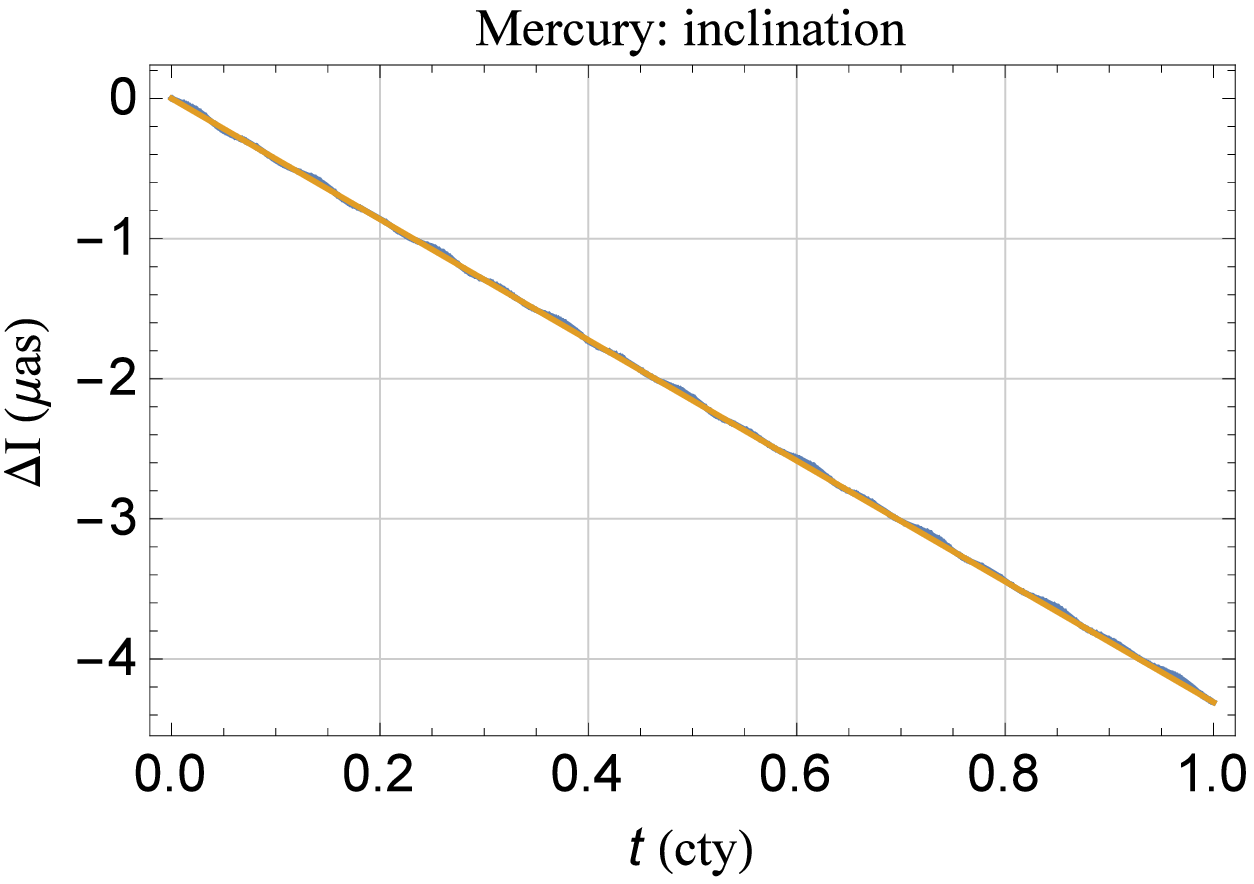}&\includegraphics[width=7cm]{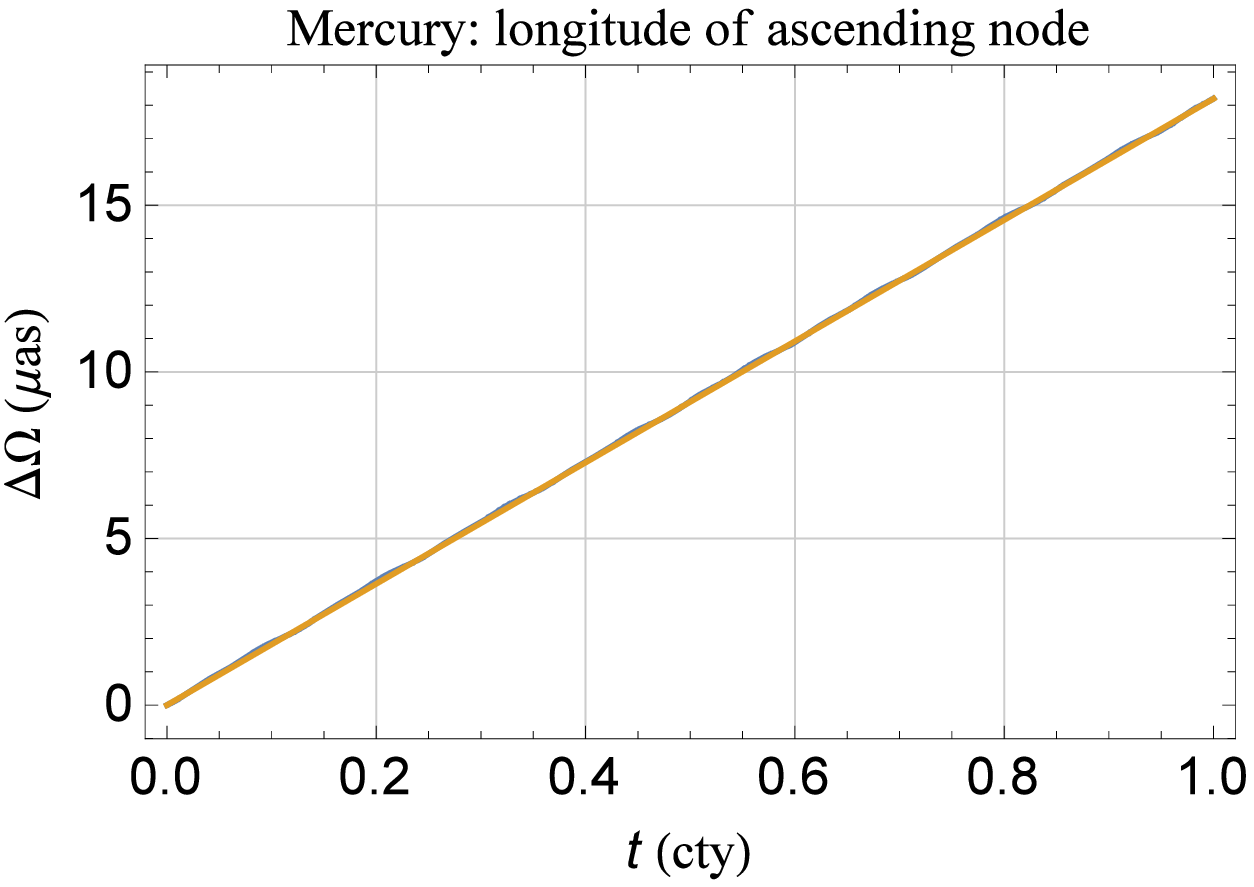}\\
\includegraphics[width=7cm]{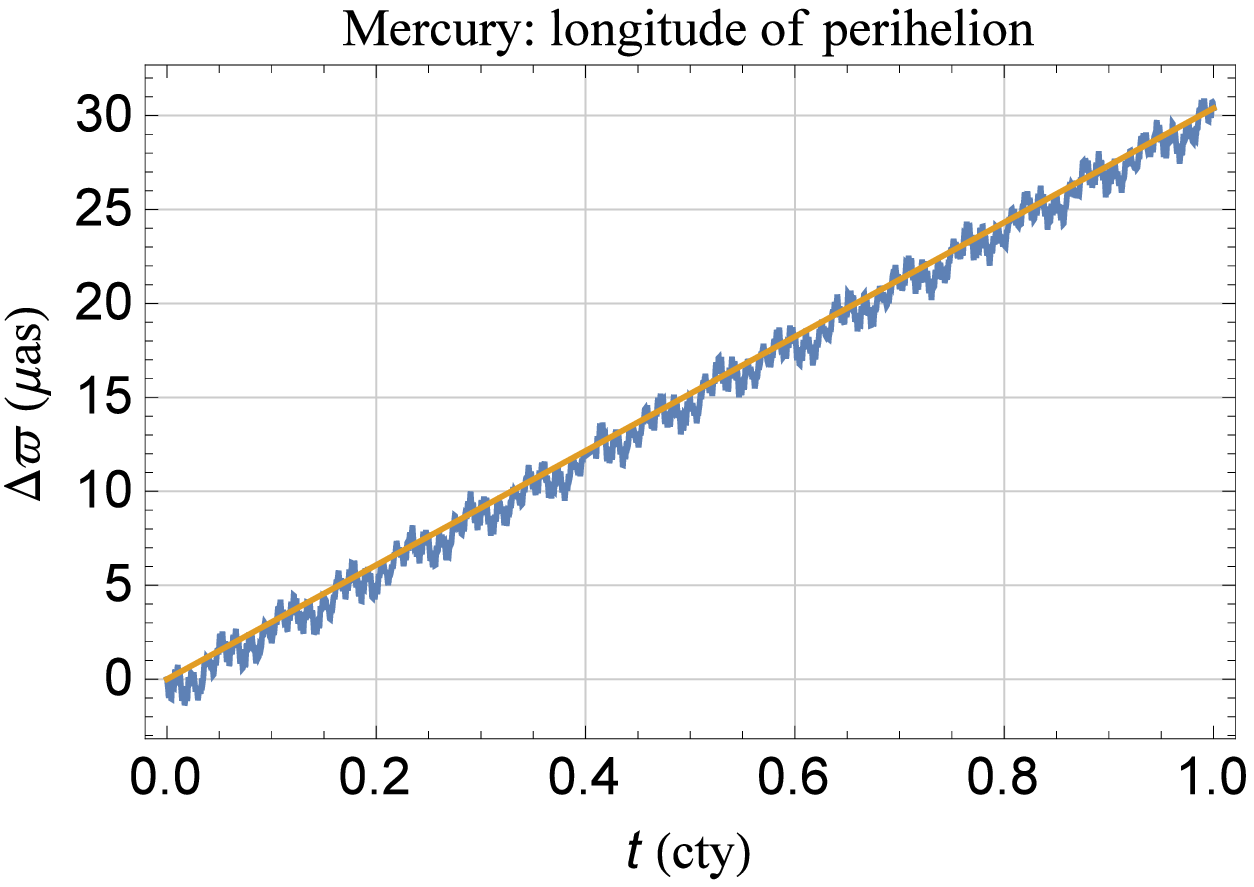}&\includegraphics[width=7cm]{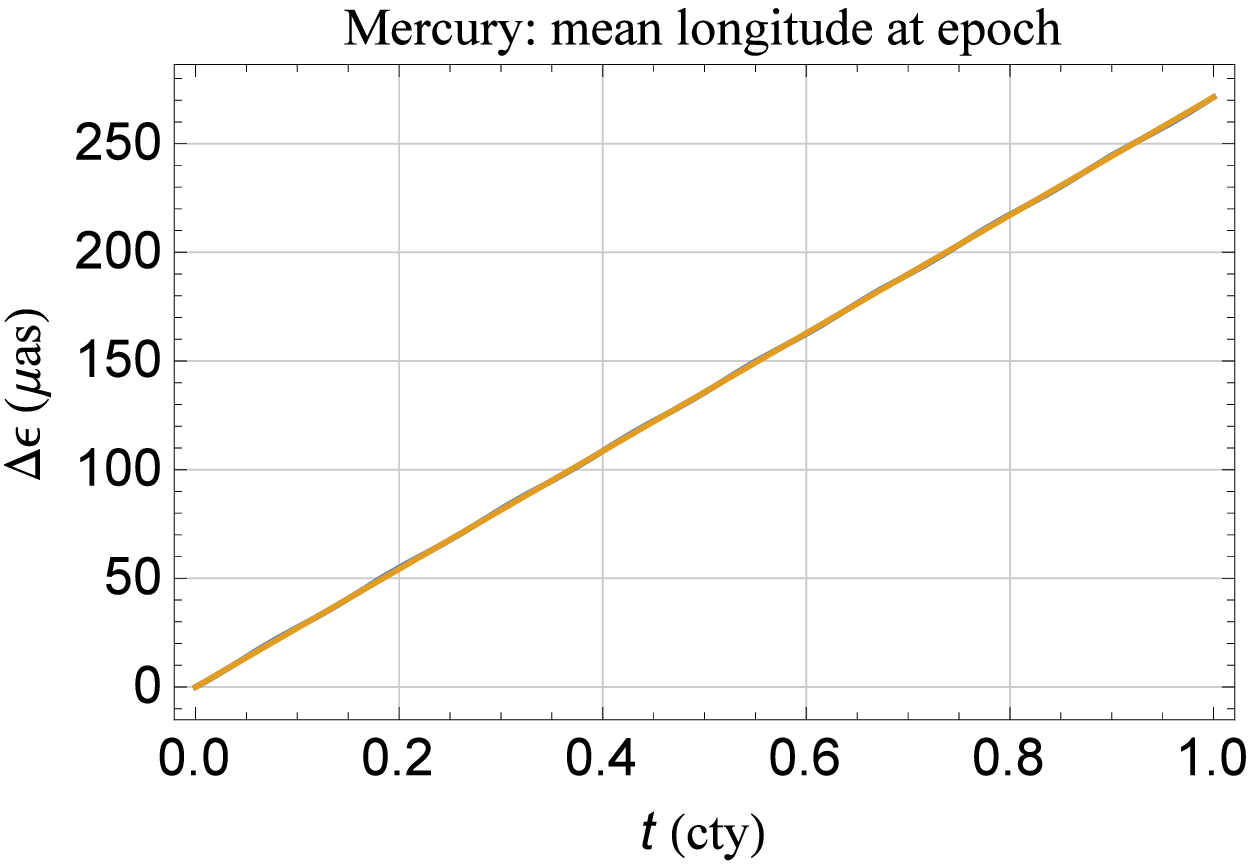}\\
\end{tabular}
\caption{Numerically integrated time series, in blue, of the shifts of the semimajor axis $a$, eccentricity $e$, inclination $I$, longitude of the ascending node $\Omega$, longitude of perihelion $\varpi$, and mean longitude at epoch $\epsilon$ of Mercury induced by the sum of all the fifteen 1pN perturbing accelerations of \rfrs{AG2}{AGM} for X ranging from Venus to Saturn over a time span 1 cty long. The units are m for $a$ and microarcseconds ($\mu$as) for all the other orbital elements. They were  obtained for each orbital element as differences between two time series calculated by numerically integrating the barycentric  equations of motion of all the planets from Mercury to Saturn in Cartesian rectangular coordinates  with and without the aforementioned  1pN $N$-body accelerations. The initial conditions, referred to the Celestial Equator at the reference epoch J2000, were retrieved from the WEB interface HORIZONS by NASA JPL; they were the same for both the integrations.  The slopes of the  secular trends, in yellow, fitted to the blue time series of $\Delta I(t),\,\Delta\Omega(t),\,\Delta\varpi(t)$, and $\Delta\epsilon(t)$ are $\dot I_\mathrm{1pN}^\mathrm{X} = -4.3\,\mu\mathrm{as\,cty}^{-1}$,  $\dot\Omega_\mathrm{1pN}^\mathrm{X} = 18.2\,\mu\mathrm{as\,cty}^{-1}$,  $\dot\varpi_\mathrm{1pN}^\mathrm{X} = 30.4\,\mu\mathrm{as\,cty}^{-1}$,  $\dot\epsilon_\mathrm{1pN}^\mathrm{X} = 271.4\,\mu\mathrm{as\,cty}^{-1}$, respectively. }\label{figura1}
\end{figure*}

\lb{numeri}
We simultaneously integrate the equations of motion of Mercury in Cartesian rectangular coordinates and the Gauss equations for each orbital element with and without the fifteen terms of the sum of \rfrs{AG2}{AGM} calculated for Venus, Earth, Mars, Jupiter and Saturn over a time span as long as 1 cty in order to clearly single out the sought features of motion: both runs share the same initial conditions retrieved on the Internet from the WEB interface HORIZONS maintained by the JPL.
For consistency reasons with the planetary data reductions available in the literature, we use the equatorial coordinates of the International Celestial Reference System (ICRS). Then, for each orbital element, we plot in Fig.\,\ref{figura1} the time series (blue curve) resulting from the difference between the runs with and without the 1pN $N-$body accelerations. Finally, we fit a linear model (yellow line) to its numerically produced signal, and estimate its slope: the outcome is collected in the caption of Fig.\,\ref{figura1}.
From Fig.\,\ref{figura1}, the secular trends of $I,\,\Omega,\,\varpi,\,\epsilon$ are apparent, while $a$ and $e$ seem to experience long-term harmonic variations. The size of the slopes of the precessions of the angular rates of change vary in the range $\simeq 1-100\,\mu\mathrm{as\,cty}^{-1}=0.001-0.1\,\mathrm{mas\,cty}^{-1}$. In particular, it turns out that the secular precession of the perihelion is about five times smaller  than \rfr{mini} \citep[Table\,2]{2018EPJC...78..549I}, being as little as
\eqi
\dot\varpi_\mathrm{1pN}^\mathrm{X}=30\,\mu\mathrm{as\,cty}^{-1} = 0.03\,\mathrm{mas\,cty}^{-1}.
\eqf
Numerical tests conducted by switching off from time to time each of \rfrs{AG2}{AGM} for every single perturbing planet X showed that the issue resides in the analytical calculation of Eq.\,(B5) in \citet{2018EPJC...78..549I} and in the consequent numerical results of the third column from the left of Table\,2 in \citet{2018EPJC...78..549I}.
\subsection{Analytical calculation}\lb{anali}
It is also possible to analytically work out the long-term rates of change of the Keplerian orbital elements of the test particle with the Gauss perturbative equations applied to \rfrs{AG2}{AGM} by doubly averaging their right-hand-sides over the orbital periods $\Pb$ and $P_\mathrm{X}$ of the perturbed body and the perturber X, respectively. The resulting expressions, especially those due to \rfrs{AG2}{AG1}, are very cumbersome. Thus, we display just approximate formulas for them to their leading order in $e$. The shifts due to \rfr{AGM}, which are relatively less involved, are displayed in full.
In the next Sects., we use the shorthand $\Delta\Omega\doteq \Omega - \Omega_\mathrm{X}$.

It turns out that there is an excellent agreement among the numerical results of Sect.\,\ref{numeri} and the analytical results  shown below.
\subsubsection{The doubly averaged rates of change of the orbital elements due to ${\bds A}_{G^2}$}\lb{analG2}
Here, we analytically calculate the doubly averaged rates of change of the Keplerian orbital elements of the test particle, to their leading order in $e$, due to \rfr{AG2}. No further approximations in the orbital configurations of both the perturbed body and X are made. They are as follows.

The semimajor axis $a$ stays constant since
\eqi
\dot a_{\bds{A}_{G^2}} \lb{G2a}  = 0.
\eqf

The rate of change of the eccentricity $e$ turns out to be
\eqi
\dot e_{\bds{A}_{G^2}} \lb{G2ecc}  = -\rp{9\,e\,\mu_\mathrm{X}\,\sqrt{\mu\,a}}{16\,c^2\,a_\mathrm{X}^3\,\ton{1-  e^2_\mathrm{X}}^{3/2}}\,\mathcal{E}_{\bds{A}_{G^2}}\ton{I,\,I_\mathrm{X},\,\Omega,\,\Omega_\mathrm{X}} + \mathcal{O}\ton{e^3},
\eqf
with
\begin{align}
\mathcal{E}_{\bds{A}_{G^2}} \nonumber \lb{globe} &= 8\,\cos 2\omega\,\ton{\cos I\,\sin^2 I_\mathrm{X}\,\sin 2\Delta\Omega - \sin I\,\sin 2I_\mathrm{X}\,\sin\Delta\Omega} -\\ \nonumber \\
\nonumber &- \sin 2\omega\,\grf{-1 + \cos 2I_\mathrm{X}\,\qua{-3 + \cos 2I\,\ton{3 + \cos 2\Delta\Omega}} - \right.\\ \nonumber \\
\nonumber & - \left. 6\,\sin^2 I_\mathrm{X}\,\cos 2\Delta\Omega  + 4\,\sin 2I\,\sin 2I_\mathrm{X}\,\cos\Delta\Omega + \right.\\ \nonumber \\
&+\left. 2\,\cos 2I\,\sin^2\Delta\Omega}.
\end{align}

As far as the rate of change of the inclination $I$ is concerned, we have
\eqi
\dot I_{\bds{A}_{G^2}} \lb{G2I}  = -\rp{3\,\mu_\mathrm{X}\,\sqrt{\mu\,a}}{c^2\,a_\mathrm{X}^3\,\ton{1-  e^2_\mathrm{X}}^{3/2}}\,\mathcal{I}_{{\bds A}_{G^2}}\ton{I,\,I_\mathrm{X},\,\Omega,\,\Omega_\mathrm{X}}+ \mathcal{O}\ton{e^2},
\eqf
with
\begin{align}
\mathcal{I}_{{\bds A}_{G^2}}&\doteq \sin I_\mathrm{X}\,\ton{\cos I\,\cos I_\mathrm{X}  + \sin I\,\sin I_\mathrm{X}\,\cos\Delta\Omega}\,\sin\Delta\Omega.
\end{align}

The precession of the node $\Omega$ is
\eqi
\dot \Omega_{\bds{A}_{G^2}} \lb{G2O}  = \rp{3\,\mu_\mathrm{X}\,\sqrt{\mu\,a}}{4\,c^2\,a_\mathrm{X}^3\,\ton{1-  e^2_\mathrm{X}}^{3/2}}\,\mathcal{N}_{\bds{A}_{G^2}}\ton{I,\,I_\mathrm{X},\,\Omega,\,\Omega_\mathrm{X}}+ \mathcal{O}\ton{e^2},
\eqf
with
\begin{align}
\mathcal{N}_{\bds{A}_{G^2}}\nonumber & -2\,\cos 2I\,\csc I\,\sin 2I_\mathrm{X}\,\cos\Delta\Omega + \\ \nonumber \\
& + \cos I\,\qua{\cos 2I_\mathrm{X}\,\ton{3 + \cos 2\Delta\Omega} + 2\,\sin^2\Delta\Omega}.
\end{align}

The precession of $\varpi$ due to \rfr{AG2} was correctly worked out, to the zero order in $e$, in Eq.\,(B2) of \citet{2018EPJC...78..549I}; thus, we do not display it here.

The rate of change of the mean longitude at epoch $\epsilon$ is
\eqi
\dot\epsilon_{\bds{A}_{G^2}} \lb{G2eps}  = \rp{\mu_\mathrm{X}\,\sqrt{\mu\,a}}{4\,c^2\,a_\mathrm{X}^3\,\ton{1-  e^2_\mathrm{X}}^{3/2}}\,\mathcal{L}_{\bds{A}_{G^2}}\ton{I,\,I_\mathrm{X},\,\Omega,\,\Omega_\mathrm{X}}+ \mathcal{O}\ton{e^2},
\eqf
where
\begin{align}
\mathcal{L}_{\bds{A}_{G^2}} \nonumber & = -1 + 3\,\cos I - 3\,\cos 2I_\mathrm{X} + 9\,\cos I\,\cos 2I_\mathrm{X} + \\ \nonumber \\
\nonumber &+ 12\,\sin^2\ton{\rp{I}{2}}\,\sin^2 I_\mathrm{X}\,\cos 2\Delta\Omega + \\ \nonumber \\
& + 6\,\ton{1 + 2\,\cos I}\,\tan\ton{\rp{I}{2}}\,\sin 2I_\mathrm{X}\,\cos\Delta\Omega.
\end{align}
\subsubsection{The doubly averaged rates of change of the orbital elements due to ${\bds A}_G$}\lb{analG1}
Here, we analytically work out the doubly averaged rates of change of the Keplerian orbital elements of the test particle, to their leading order in $e$, induced by \rfr{AG1}. No further approximations in the orbital configurations of both the perturbed body and X are made. We list them below.

For the semimajor axis $a$, we have
\eqi
\dot a_{\bds{A}_{G}} \lb{esattasemia}  = \rp{3\,\mu_\mathrm{X}\,a^{3/2}\,\sqrt{\mu}}{2\,c^2\,a_\mathrm{X}^3\,\ton{1-  e^2_\mathrm{X}}^{3/2}}\,\mathcal{A}_{\bds{A}_{G}}\ton{I,\,I_\mathrm{X},\,\Omega,\,\Omega_\mathrm{X}} + \mathcal{O}\ton{e^2},
\eqf
with
\eqi
\mathcal{A}_{\bds{A}_{G}} = \sin I_\mathrm{X}\,\ton{-\sin I\,\cos I_\mathrm{X}+ \cos I\,\sin I_\mathrm{X}\,\cos\Delta\Omega}\,\sin\Delta\Omega.
\eqf

The rate of change of the eccentricity $e$ is
\eqi
\dot e_{\bds{A}_{G}} \lb{esattaecc}  = -\rp{3\,e\,\mu_\mathrm{X}\,\sqrt{\mu\,a}}{2\,c^2\,a_\mathrm{X}^3\,\ton{1-  e^2_\mathrm{X}}^{3/2}}\,\mathcal{E}_{\bds{A}_{G}}\ton{I,\,I_\mathrm{X},\,\Omega,\,\Omega_\mathrm{X}} + \mathcal{O}\ton{e^3},
\eqf
with
\eqi
\mathcal{E}_{\bds{A}_{G}} = \sin I_\mathrm{X}\,\ton{-\sin I\,\cos I_\mathrm{X}+ \cos I\,\sin I_\mathrm{X}\,\cos\Delta\Omega}\,\sin\Delta\Omega.
\eqf

The rate of change of the inclination $I$ turns out to be
\eqi
\dot I_{\bds{A}_{G}} \lb{esattai}  = \rp{3\,\mu_\mathrm{X}\,\sqrt{\mu\,a}}{4\,c^2\,a_\mathrm{X}^3\,\ton{1-  e^2_\mathrm{X}}^{3/2}}\,\mathcal{I}_{\bds{A}_{G}}\ton{I,\,I_\mathrm{X},\,\Omega,\,\Omega_\mathrm{X}}+ \mathcal{O}\ton{e^2},
\eqf
with
\begin{align}
\mathcal{I}_{\bds{A}_{G}}&\doteq \sin I_\mathrm{X}\,\ton{\cos I\,\cos I_\mathrm{X}  + \sin I\,\sin I_\mathrm{X}\,\cos\Delta\Omega}\,\sin\Delta\Omega.
\end{align}

The precession of the node $\Omega$ is
\eqi
\dot \Omega_{\bds{A}_{G}} \lb{esattaO}  = -\rp{3\,\mu_\mathrm{X}\,\sqrt{\mu\,a}}{16\,c^2\,a_\mathrm{X}^3\,\ton{1-  e^2_\mathrm{X}}^{3/2}}\,\mathcal{N}_{\bds{A}_{G}}\ton{I,\,I_\mathrm{X},\,\Omega,\,\Omega_\mathrm{X}}+ \mathcal{O}\ton{e^2},
\eqf
with
\begin{align}
\mathcal{N}_{\bds{A}_{G}}\nonumber &\doteq -2\,\cos 2I\,\csc I\,\sin 2I_\mathrm{X}\,\cos\Delta\Omega + \\ \nonumber \\
& + \cos I\,\qua{\cos 2I_\mathrm{X}\,\ton{3 + \cos 2\Delta\Omega} + 2\,\sin^2\Delta\Omega}.
\end{align}

For the precession of the longitude of perihelion $\varpi$, we have
\eqi
\dot\varpi_{\bds{A}_{G}} \lb{esattaw}  = -\rp{\mu_\mathrm{X}\,\sqrt{\mu\,a}\,\csc I}{8\,c^2\,a_\mathrm{X}^3\,\ton{1-  e^2_\mathrm{X}}^{3/2}}\,\mathcal{W}\ton{I,\,I_\mathrm{X},\,\Omega,\,\Omega_\mathrm{X}}+ \mathcal{O}\ton{e^2},
\eqf
with
\begin{align}
\mathcal{W}\nonumber \lb{doppiavu}&\doteq \rp{9}{2}\sin^3 I\,
\qua{-2+\sin^2 I_\mathrm{X}\,\ton{3+\cos 2\Delta\Omega} }+\\ \nonumber \\
\nonumber &+\sin I\,\grf{2 + 6\,\cos I + 6\cos 2 I_\mathrm{X} - \right.\\ \nonumber \\
\nonumber & -\left. 3\,\sin^2 I_\mathrm{X}\,\qua{3\,\cos I + \ton{-3+ \cos I }\cos 2\Delta\Omega }} -\\ \nonumber \\
\nonumber &- 6\,\sin^2\ton{\rp{I}{2}}\,\sin 2I_\mathrm{X}\,\cos\Delta\Omega + \\ \nonumber \\
& +3\,\ton{2 + 3\,\cos I}\,\sin^2 I\,\sin 2I_\mathrm{X}\,\cos\Delta\Omega.
\end{align}
\Rfrs{esattaw}{doppiavu}, which correct Eq.\,(B5) of \citet{2018EPJC...78..549I}, allow to calculate the same values for Mercury which are obtained with our numerical integrations of Sect.\,\ref{numeri}, limited to \rfr{AG1} only, for each of the perturbing planets at a time.

The rate of change of the mean longitude at epoch $\epsilon$ is given by
\eqi
\dot\epsilon_{\bds{A}_{G}} \lb{esattaeps}  = -\rp{\mu_\mathrm{X}\,\sqrt{\mu\,a}}{32\,c^2\,a_\mathrm{X}^3\,\ton{1-  e^2_\mathrm{X}}^{3/2}}\,\mathcal{L}_{\bds{A}_{G}}\ton{I,\,I_\mathrm{X},\,\Omega,\,\Omega_\mathrm{X}}+ \mathcal{O}\ton{e^2},
\eqf
with
\begin{align}
\mathcal{L}_{\bds{A}_{G}} \nonumber & = \ton{-1 + 6\,\cos I + 3\,\cos 2 I}\,\ton{1 + 3\,\cos 2 I_\mathrm{X}} + \\ \nonumber \\
\nonumber & +  24\,\ton{2 + \cos I}\,\sin^2 \ton{\rp{I}{2}}\,\sin^2 I_\mathrm{X}\cos 2 \Omega\,\cos 2 \Omega_\mathrm{X} + \\ \nonumber \\
\nonumber & + 6\,\sec \ton{\rp{I}{2}}\,\qua{3\,\sin \ton{3\,\rp{I}{2}} + \sin \ton{5\,\rp{I}{2}}}\,\sin 2 I_\mathrm{X}\cos \Omega\,\cos \Omega_\mathrm{X} +\\ \nonumber \\
\nonumber & +  6\,\sec \ton{\rp{I}{2}}\,\qua{3 \sin \ton{3\,\rp{I}{2}} + \sin \ton{5\,\rp{I}{2}}}\,\sin 2 I_\mathrm{X}\sin \Omega\,\sin \Omega_\mathrm{X} +\\ \nonumber \\
& + 24\,\ton{2 + \cos I}\,\sin^2  \ton{\rp{I}{2}}\,\sin^2 I_\mathrm{X}\sin 2 \Omega\,\sin 2 \Omega_\mathrm{X}.
\end{align}
\subsubsection{The doubly averaged rates of change of the orbital elements due to ${\bds A}_{{\bds v}_\mathrm{X}}$}\lb{analGM}
Here, we analytically calculate the doubly averaged rates of change of the Keplerian orbital elements of the test particle caused by \rfr{AGM}. No approximations in the orbital configurations of both the perturbed body and X are made; the following expressions are exact.

The semimajor axis $a$ and the eccentricity $e$  are constant since
\begin{align}
\dot a_{\bds{A}_{{\bds v}_\mathrm{X}}} \lb{GMa}  &= 0, \\ \nonumber \\
\dot e_{\bds{A}_{{\bds v}_\mathrm{X}}} \lb{GMe}  &= 0.
\end{align}

The rate of change of the inclination $I$ is
\eqi
\dot I_{\bds{A}_{{\bds v}_\mathrm{X}}} \lb{GMI}  = -\rp{2\,\mu_\mathrm{X}\,\sqrt{\mu}\,\sin I_\mathrm{X}\,\sin\Delta\Omega}{c^2\,a_\mathrm{X}^{5/2}\,\ton{1-e^2_\mathrm{X}}}.
\eqf

For the precession of the node $\Omega$ we have
\eqi
\dot \Omega_{\bds{A}_{{\bds v}_\mathrm{X}}} \lb{GMO}  = \rp{2\,\mu_\mathrm{X}\,\sqrt{\mu}\,\ton{\cos I_\mathrm{X} - \cot I\,\sin I_\mathrm{X}\,\cos\Delta\Omega}}{c^2\,a_\mathrm{X}^{5/2}\,\ton{1-e^2_\mathrm{X}}}.
\eqf

The precession of $\varpi$ due to \rfr{AGM} was correctly calculated in Eq.\,(B8) of \citet{2018EPJC...78..549I}; as such, it is not shown here.

The rate of change of the mean longitude at epoch $\epsilon$ does depend on $e$. It turns out to be
\eqi
\dot \epsilon_{\bds{A}_{{\bds v}_\mathrm{X}}} \lb{GMeps}  = \rp{2\,\mu_\mathrm{X}\,\sqrt{\mu}}{c^2\,a_\mathrm{X}^{5/2}\,\ton{1-e^2_\mathrm{X}}}\mathcal{L}_{{\bds v}_\mathrm{X}}\ton{I,\,\Omega,\,I_\mathrm{X},\,\Omega_\mathrm{X}},
\eqf
where
\begin{align}
\mathcal{L}_{{\bds v}_\mathrm{X}} \nonumber  & = \ton{1 + 3\,\sqrt{1 - e^2}\,\cos I}\,\cos I_\mathrm{X} + \\ \nonumber \\
& + \ton{1 + 3\,\sqrt{1 - e^2} + 3\,\sqrt{1 - e^2}\,\cos I}\,\tan\ton{\rp{I}{2}}\,\sin I_\mathrm{X}\,\cos\Delta\Omega.
\end{align}
\section{Confrontation with the observations}\lb{ossser}
\citet{2019AJ....157..220I} attempted to calculate the formal uncertainties in the secular rates of change of $a,\,e,\,I,\,\Omega$, and $\varpi$ of the planets of the solar system from the recently released formal errors in $a$ and the nonsingular orbital elements $e\sin\varpi,\,e\cos\varpi,\,\sin I\sin\Omega$, and $\sin I\cos\Omega$ estimated for the same bodies with the EPM2017 ephemerides by \citet{2018AstL...44..554P}. Since, among other things, the 1pN $N$-body equations of motion are routinely included in the EPM software dynamics, such errors should be overall  regarded as representative of the current level of modeling the solar system dynamics along with measurement errors. As such, they may be viewed as the uncertainties that would affect a putative measurement of the effects worked out in Sect.\,\ref{calcoli} if they were explicitly measured in some dedicated data analysis.
From the column dedicated to Mercury in Table\,1 of \citet{2019AJ....157..220I}, it can be noted that the $1-\upsigma$ error in $\dot a$ amounts to $\delta\dot a_\mathrm{obs} = 0.003\,\mathrm{m\,cty}^{-1}$, while for the other Keplerian orbital elements we have $\delta\dot e_\mathrm{obs} = 0.6\,\mu\mathrm{as\,cty}^{-1},\,\delta\dot I_\mathrm{obs} = 3\,\mu\mathrm{as\,cty}^{-1},\,\delta\dot\Omega_\mathrm{obs}=24\,\mu\mathrm{as\,cty}^{-1},$ and $\delta\dot\varpi_\mathrm{obs} = 8\,\mu\mathrm{as\,cty}^{-1}$.
From a comparison with the expected 1pN rates of change of Fig.\,\ref{figura1}, it turns out that, with the possible exception of the perihelion, they are about of the same order of magnitude of the aforementioned uncertainties. Moreover, as discussed in \citet{2018AstL...44..554P} and \citet{2019AJ....157..220I}, the latter ones may be optimistic.
Thus, it is difficult to deem the predicted 1pN $N$-body precession $\dot\varpi_\mathrm{1pN}^\mathrm{X} = 30\,\mu\mathrm{as\,cty}^{-1}$ as realistically measurable compared to a merely formal uncertainty $\delta\dot\varpi_\mathrm{obs} = 8\,\mu\mathrm{as\,cty}^{-1}$. It is worth noticing that such a tiny error would correspond to current bounds in the PPN parameters $\beta,\,\gamma$ as little as $\simeq 10^{-7}$, which are better than the expected accuracy from the ongoing BepiColombo mission quoted by \citet{2018PhRvL.120s1101W}; see the discussion in \citet{2019AJ....157..220I} about the reliability of such an evaluation.
The mean longitude at epoch $\epsilon$ seem, at first sight, more interesting since its 1pN $N$-body rate is as large as $\dot\epsilon^\mathrm{X}_\mathrm{1pN} = 270\,\mu\mathrm{as\,cty}^{-1}=0.27\,\mathrm{mas\,cty^{-1}}$.
\citet{2019AJ....157..220I} did not calculate the uncertainty in $\dot\epsilon$. In their Table\,3, \citet{2018AstL...44..554P} released the formal uncertainty in the planetary  mean longitudes, dubbed there as $\lambda$; for Mercury, it is as little as $\delta\lambda_\mathrm{obs} = 3.3\,\mu\mathrm{as}$. This implies that, in order to retrieve the uncertainty in $\dot\epsilon$, the errors in the mean motion $\nk$ due to the mismodeling of the Sun's gravitational parameter $\mu$ and of the planet's semimajor axis are required as well. Since $\delta\mu_\mathrm{obs}=1\times 10^{10}\,\mathrm{m^3\,s^{-2}}$ \citep{2015JPCRD..44c1210P}, the resulting error in the Hermean mean motion is as large as $\delta\nk^\mathrm{obs} = 20\,\mathrm{mas\,cty}^{-1}$. It vanishes the possibility of measuring the 1pN $N$-body effect on $\epsilon$.
As such, only a dramatic improvement in the determination of the Hermean orbit, which might be obtained when all the data from BepiColombo will be collected and processed, may bring the 1pN $N$-body precessions due to the direct effect of \rfrs{AG2}{AGM} in the measurability domain.

On the other hand, even should this finally be the case, the concerns raised by \citet{2018EPJC...78..549I} about the systematic errors caused by the competing Sun's quadrupole and Lense-Thirring rates of change are even reinforced by the present analysis since the actual size of the 1pN $N$-body perihelion precession of Mercury turned out to be smaller than the incorrect value of \rfr{mini}. Thus, it is hopeful that the astronomers will finally provide the community with the supplementary advances of all the other Keplerian orbital elements in addition to the perihelion. Indeed, if and when it will happen, it would, then, be possible to set up linear combinations of them suitably designed to cancel out, by construction, the other unwanted precessions. An analogous approach, originally limited just to the perihelia of other planets and asteroids in order to separate the disturbing Sun's $J_2$ action from the Schwar\textcolor{black}{z}schild-type rates of changes was proposed by \citet{1990grg..conf..313S}. It is also widely used in ongoing relativistic tests with geodetic satellites in the Earth's field; see, e.g., \citet{2013CEJPh..11..531R}, and references therein for an overview.
\section{Summary and conclusions}\lb{conclu}
Recently, \citet{2018PhRvL.120s1101W} calculated a new general relativistic contribution to the Mercury's perihelion advance as large as $\dot\varpi_\mathrm{1pN}^\mathrm{X} = 220\,\mu\mathrm{as\,cty}^{-1}$ arising from an approximated form of the 1pN $N$-body equations of motion restricted to a \textcolor{black}{hierarchical} three body system. He claimed that it may be measured in the next future by the ongoing BepiColombo mission to Mercury if it will reach a $\simeq 10^{-6}$ accuracy level in constraining the PPN parameters $\beta,\,\gamma$. Later, the present author first remarked in \citet{2018EPJC...78..549I} that the indirect precession due to the interplay of the Newtonian $N$-body and the 1pN Sun's Schwarzschild-like accelerations in the equations of motion is likely undetectable in actual data reductions since it cannot be expressed in terms of a dedicated, solve-for parameter scaling an acceleration different from the aforementioned ones which are routinely modeled. Then, he calculated analytically the individual contributions to the perihelion advance induced directly by each of the approximated 1pN $N$-body accelerations put forth by \citet{2018PhRvL.120s1101W} by finding an overall precession of  $\dot\varpi_\mathrm{1pN}^\mathrm{X} = 150\,\mu\mathrm{as\,cty}^{-1}$. \citet{2018EPJC...78..549I} discussed also the impact of the systematic aliasing due to the competing perihelion rates induced by the Sun's quadrupole mass moment $J_2$ and angular momentum $\bds{A}$ via the Lense-Thirring effect by noting that their mismodeling would likely compromise a clean recovery of the 1pN effect of interest.

Here, the secular rates of change of all the other Keplerian orbital elements $a,\,e,\,I,\,\Omega,\,\varpi$, and $\epsilon$ caused by the same approximated 1pN $N$-body accelerations by \citet{2018PhRvL.120s1101W} were analytically worked out. A numerical integration of the equations of motion confirmed such findings in the case of Mercury acted upon by the other planets from Venus to Saturn. The resulting rates of change amount to $\dot I_\mathrm{1pN}^\mathrm{X} = -4.3\,\ton{\mu\mathrm{as\,cty}^{-1}}$,  $\dot\Omega_\mathrm{1pN}^\mathrm{X} = 18.2\,\mu\mathrm{as\,cty}^{-1}$,  $\dot\varpi_\mathrm{1pN}^\mathrm{X} = 30.4\,\mu\mathrm{as\,cty}^{-1}$,  $\dot\epsilon_\mathrm{1pN}^\mathrm{X} = 271.4\,\mu\mathrm{as\,cty}^{-1}$. As a result, the Hermean 1pN $N$-body perihelion precession turned out to be smaller than the previously reported values because of an error explicitly disclosed, at least in the calculation by \citet{2018EPJC...78..549I}. This makes even more difficult than before its possible present and future measurement. A comparison with the merely formal uncertainties in some of the orbital secular rates of Mercury, recently obtained by \citet{2019AJ....157..220I} from the EPM2017 ephemerides, showed that the sizes of the predicted 1pN $N$-body precessions are just at the same level or even below them if, more realistically, they are rescaled by a factor of $\simeq 10-50$ \citep{2019AJ....157..220I}. If our future knowledge of the orbit of the closest planet to the Sun will be adequately improved, the systematic bias caused by other competing precessions could be removed by suitably designing linear combinations of the other Keplerian orbital elements of Mercury, provided that the astronomers will determine also their supplementary advances in addition to the perihelion's one.
\bibliography{MR_biblio,MS_binary_pulsar_bib,Gclockbib,semimabib,PXbib}{}


\end{document}